**Coexistence of negative and positive photoconductivity in few-layer PtSe2 field-effect transistors**

*Alessandro Grillo, Enver Faella, Aniello Pelella, Filippo Giubileo, Lida Ansari, Farzan Gity, Paul K. Hurley, Niall McEvoy, and Antonio Di Bartolomeo\**

Alessandro Grillo, Enver Faella, Aniello Pelella, Prof. Antonio Di Bartolomeo
Physics Department "E. R. Caianiello", University of Salerno, via Giovanni Paolo II n. 132, Fisciano 84084, Salerno, Italy
E-mail: adibartolomeo@unisa.it

Dr. Filippo Giubileo
CNR-SPIN Salerno, via Giovanni Paolo II n. 132, Fisciano 84084, Salerno, Italy

Dr. Lida Ansari, Dr. Farzan Gity, Prof. Paul K. Hurley
Tyndall National Institute, University College Cork, Cork, Ireland

Prof. Niall McEvoy
AMBER & School of Chemistry, Trinity College Dublin, Dublin 2, Ireland

Platinum diselenide ($PtSe_2$) field-effect transistors with ultrathin channel regions exhibit p-type electrical conductivity that is sensitive to temperature and environmental pressure. Exposure to a supercontinuum white light source reveals that positive and negative photoconductivity coexists in the same device. The dominance of one type of photoconductivity over the other is controlled by environmental pressure. Indeed, positive photoconductivity observed in high vacuum converts to negative photoconductivity when the pressure is rised. Density functional theory calculations confirm that physisorbed oxygen molecules on the $PtSe_2$ surface act as acceptors. The desorption of oxygen molecules from the surface, caused by light irradiation, leads to decreased carrier concentration in the channel conductivity. The understanding of the charge transfer occurring between the physisorbed oxygen molecules and the $PtSe_2$ film provides an effective route for modulating the density of carriers and the optical properties of the material.

# 1. Introduction

Two-dimensional (2D) transition metal dichalcogenides (TMDs)[1,2] have demonstrated great potential in applications such as field-effect transistors,[3] photodetectors,[4] solar cells,[5] field emitters,[6,7] memory devices,[8] and gas sensors.[9,10] Noble TMDs like platinum diselenide, $PtSe_2$, are currently the subject of intense research endeavour.[11–13] Bulk $PtSe_2$ is a gapless semimetal[14] for which theoretical studies have predicted a band gap emergence when reduced to monolayer.[15] Wang et. Al. epitaxially grew the material by direct selenization and experimentally proved that monolayer $PtSe_2$ has a band gap of $1.2\ eV$.[16] Owing to its thickness-dependent semimetal-to-semiconductor transition, $PtSe_2$ is a suitable candidate for electronic[17] and optoelectronic applications.[18] Nowadays, $PtSe_2$ is attracting great research attentions due to its high carrier mobility[13], high chemical activity,[19] and low formation temperature.[20] Yim et al. demonstrated that $PtSe_2$ films, transferred onto silicon substrates, form diodes working as both photodetectors and photovoltaic cells.[21] Moreover, theoretical studies have indicated that $PtSe_2$ has adsorption energy for gases like $NO_2$ and $NH_3$ lower than $MoS_2$ and graphene.[22] Su et al. have recently reported stable gas sensing performance of $PtSe_2$ under strains induced by different selenization temperatures.[23] $PtSe_2$ has weak chemical reactivity with oxygen that is physically adsorbed on its surface and can be removed by vacuum annealing, heating, or irradiation.[24] The interaction of $O_2$ with $PtSe_2$ is an important process to understand because it can deeply affect the transport properties of the material. The adsorption process does not lead to structural or chemical changes of the material but induces p-type doping as electrons can transfer to oxygen molecules and cause hole formation inside $PtSe_2$.

In this work, we investigate the effect of temperature, air pressure, and light irradiation on the transport properties of multilayer $PtSe_2$ back-gated field-effect transistors. The devices show p-type conduction and electrical conductivity that increases with the rising temperature and air pressure. We demonstrate that negative and positive photoconductivity coexist in the device, and their relative contributions depend on the air pressure. Negative photoconductivity (NPC) is observed at atmospheric pressure as light irradiation by a white source reduces the $PtSe_2$ conductance. Conversely, in high vacuum, the material exhibits positive photoconductivity (PPC). The physical origin of this mechanisms is the pressure- or light-induced desorption of oxygen, which acts as a charge transfer dopant, affecting the $PtSe_2$ conductance. To provide atomistic insight, we study the interaction of physisorbed oxygen molecules with semiconducting $PtSe_2$ film by performing first-principles calculations, demonstrating that adsorption of oxygen on $PtSe_2$ surface induces holes in the $PtSe_2$ material, increasing its conductivity, consistently with the experimental observations.

## 2. Results and discussion

$PtSe_2$ has a 1T crystal structure with octahedral coordination (in which six selenium atoms are bonded to a platinum atom located at the centre, as shown in **Figure 1a**), resulting in a $Pt$ layer sandwiched between Se layers.[25] **Figure 1b** shows a transmission electron microscope (TEM) image of a $PtSe_2$ flake used for the electrical characterization. It reveals that the thickness is about 3 nm corresponding to 6 atomic layers.[26] Details about the fabrication are provided in the materials and methods section.

**Figure 1c** and **1d** shows the electrical characterization of the $PtSe_2$ back-gated field-effect transistor at room temperature. The output characteristics, i.e. the drain current ($I_d$) as a function of the voltage between the two innermost contacts ($V_{ds}$), with the gate-source voltage ($V_{gs}$) as control parameter, exhibit a linear behaviour (Figure 1c). The application of the gate voltage affects the overall conductance of the transistor without modifying the $I_d - V_{ds}$ linearity, with $I_d$ reduced by increasing $V_{gs}$.

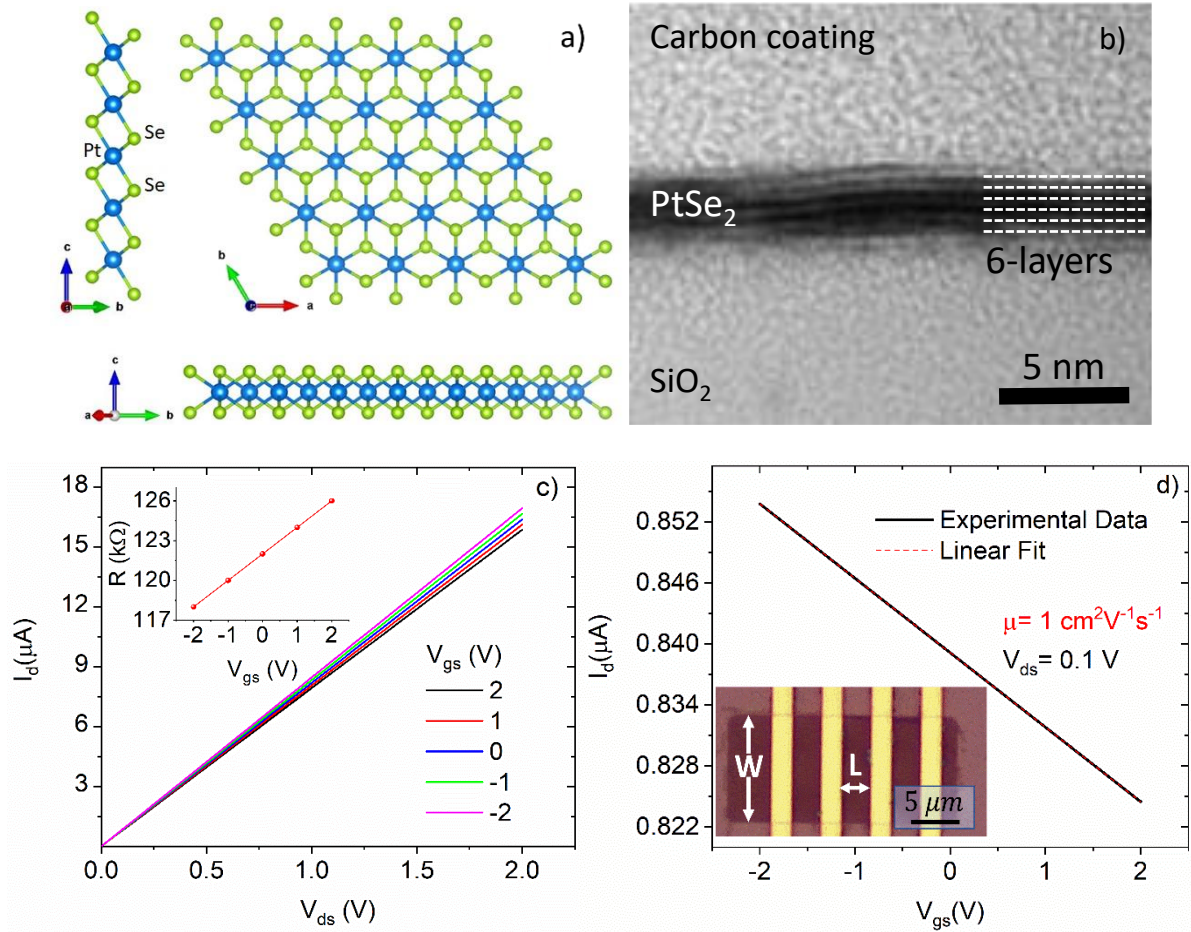

*Figure 1 – PtSe₂ structure and transistor characterization a) Crystal structure of two-dimensional $PtSe_2$ showing an octahedral geometry with six selenium atoms bonded to a platinum atom located at the centre. b) TEM image of a $PtSe_2$ film revealing a thickness of about $3\ nm$ corresponding to 6 layers. c) Output characteristics acquired in 2-probe configuration for different gate voltages. The inset shows the channel resistance as function of the gate voltage. d) Transfer characteristic recorded at $V_{ds} = 0.1\ V$. The dashed red line shows the linear fit used to obtain the field-effect mobility. The inset shows an optical image of the device. W and L represent the width and the length of the selected channel, respectively.*

The transfer characteristic, i.e. the $I_d - V_{gs}$ curve measured over a gate voltage sweep from $-2V$ to $+2V$ is reported in Figure 1d. It shows a typical p-type behaviour with higher conductance at negative $V_{gs}$. The gate voltage range was intentionally limited to avoid leakage or breakdown of the $85\ nm$ gate oxide. The intrinsic p-type doping has been reported in previous works and can be attributed to $Pt$ or $Se$ vacancies and to the effect of oxygen adsorbates on the channel surface.[17,24,27] Furthermore, the use of $Ni$ as the contact material facilitates hole injection as the $Ni$ Fermi level aligns to the top of the valence band of $PtSe_2$. The field-effect mobility, defined as $\mu_{FE} = \frac{L}{WC_{ox}V_{ds}}\frac{dI_d}{dV_{gs}}$, has been obtained through the linear fit of the transfer curve ($L = 4\ \mu m$ and $W = 10\ \mu m$ are the channel length and width, respectively, $C_{ox} = \frac{\epsilon_0 \cdot \epsilon_{SiO_2}}{t_{SiO_2}} = 4.06 \cdot 10^{-8}\ \frac{F}{cm^2}$ is the capacitance per unit area of the gate dielectric with $\epsilon_0 = 8.85 \cdot 10^{-14}\ \frac{F}{cm}$, $\epsilon_{SiO_2} = 3.9$ and $t_{SiO_2} = 85\ nm$ the vacuum permittivity, the $SiO_2$ relative permittivity and thickness, respectively; $V_{ds} = 0.1\ V$ is the voltage bias between drain and source). The obtained $\mu_{FE} \approx 1\ cm^2V^{-1}s^{-1}$ is comparable with the one reported for differently fabricated $PtSe_2$ devices.[27,28]

The temperature and pressure dependence of the electrical characteristics of the device was investigated by measuring the output and transfer characteristics in the $292 - 392\ K$ temperature range and at different air pressures of $10^3$, $1$ and $10^{-5}\ mbar$, as reported in the Supporting Information. We found an increasing current/conductance with both rising temperature and air pressure, without any other apparent change in the behaviour of the device. **Figure 2a** shows that the conductance $G$ increases as the temperature rises, confirming a semiconducting behaviour. The air pressure also affects the conductance of the device. Indeed, the device is more conductive at atmospheric pressure than in vacuum. This effect is caused by oxygen that acts as a p-type dopant for the material,[29] and it will be further discussed in the next section examining the impact of physisorbed oxygen on the $PtSe_2$ electronic properties through DFT calculations.

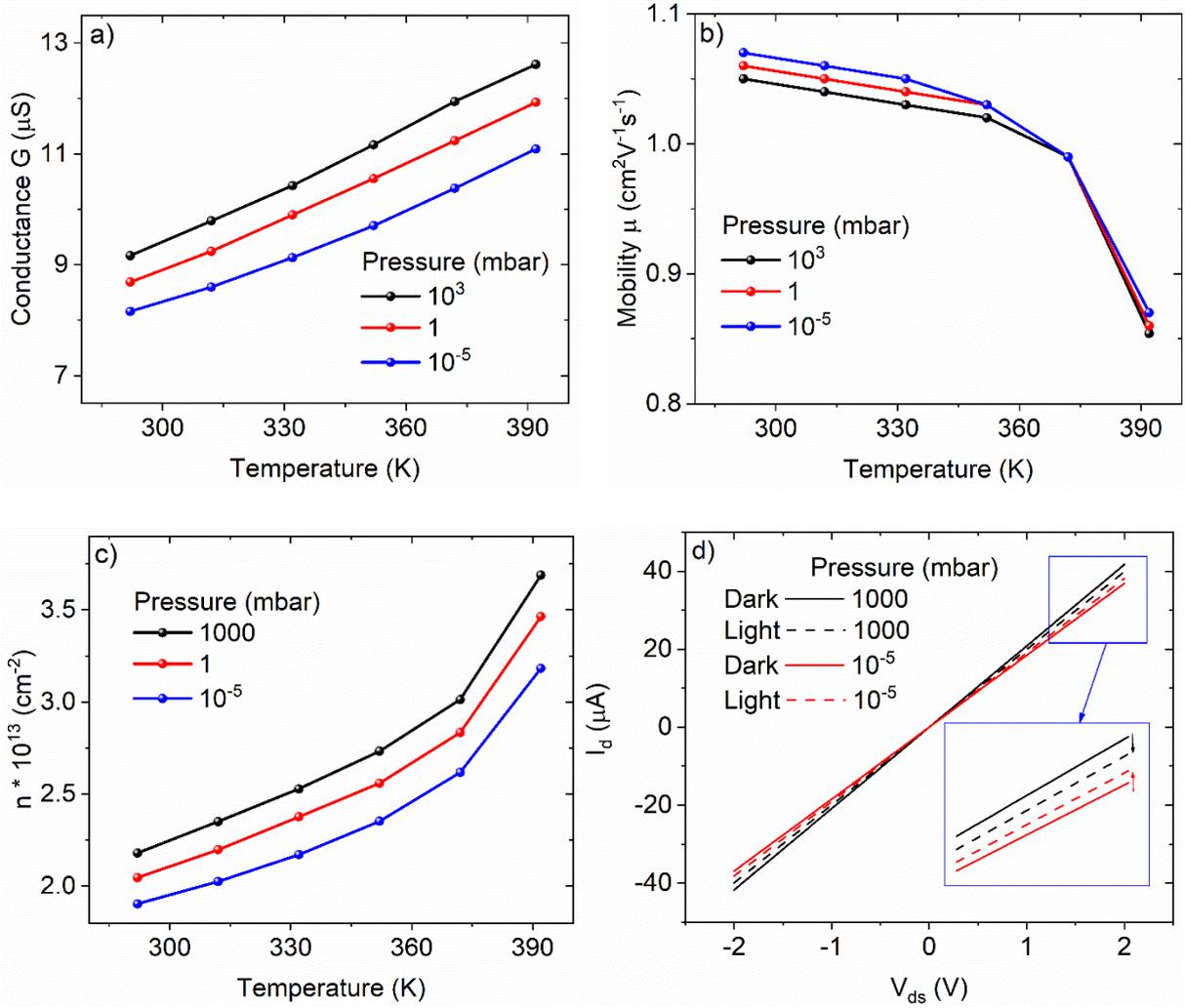

*Figure 2 – **Temperature dependent transistor parameters.** a) Channel conductance, b) field-effect mobility and c) carrier density as a function of temperature for $1\,bar$ (black dots), $1\,mbar$ (red dots), $10^{-5}\,mbar$ (blue dots). d) $I_d - V_{ds}$ characteristics in dark (continuous lines) and under super-continuous 450-2400 nm light source irradiation (dashed lines), recorded at $1\,bar$ (black lines) and $10^{-5}\,mbar$ (black lines).*

**Figure 2b** shows that the mobility depends weakly on environmental pressure and temperature below 350 K.[30] Usually, in a semiconductor the mobility is affected by two competitive mechanisms, i.e. ionized impurity scattering and phonon scattering.[31] The first mechanism dominates at lower temperatures and yields an increasing mobility with rising temperature, while phonon scattering becomes the prevailing mechanism at higher temperatures and causes a decreasing mobility. In the device under study, the two phenomena balance each other with a slight predominance of phonon scattering which becomes relevant when the temperature rises beyond 350 K. Having obtained $\mu_{FE}(T)$ and $G(T)$ at each of the considered pressures, we estimate the carrier density per unit area $n$ (in $cm^{-2}$) as a function of temperature for each pressure from the relation $n(T) = \frac{G(T)}{e\mu_{FE}(T)}\frac{L}{W}$, where $e$ is the electron charge. **Figure 2c** shows the typical semiconductor behaviour where the carrier density increases with the temperature. The obtained carrier concentration of about $10^{13}\ cm^{-2}$ is comparable with the one reported from Hall measurements carried out on similar devices.[17] Moreover, we notice that at atmospheric pressure the carrier density is larger than in high vacuum. This could be due to enhanced adsorption of oxygen on the channel surface at higher pressure that leads to increased p-type doping.

**Figure 2d** shows $I_d - V_{ds}$ characteristics recorded at air pressures of $1\ bar$ and $10^{-5}\ mbar$ both in dark and under white light irradiation. In high vacuum the light irradiation causes an increase in the conductivity as expected for a semiconducting material. Conversely, at atmospheric pressure, a reduction in conductivity, i.e. a negative photoconductivity, is observed under illumination. We remark that the linearity of the $I_d - V_{ds}$ characteristics of Figure 2d indicates that the drop in carrier concentration reported in Figure 2c is not caused by $Ni/PtSe_2$ contacts becoming non-ohmic; otherwise stated, the current is not reduced by the formation of a Schottky barrier at metal/$PtSe_2$ interface but rather by a decrease in channel conductivity.

To further investigate the optoelectronic behaviour of $PtSe_2$, we measured the electrical conduction of the device under light irradiation at different air pressures. A schematic of the experimental setup is reported in **Figure 3a**. The super-continuous light source power was set to $20\ mW/mm^2$ and switched on and off every 3 minutes.

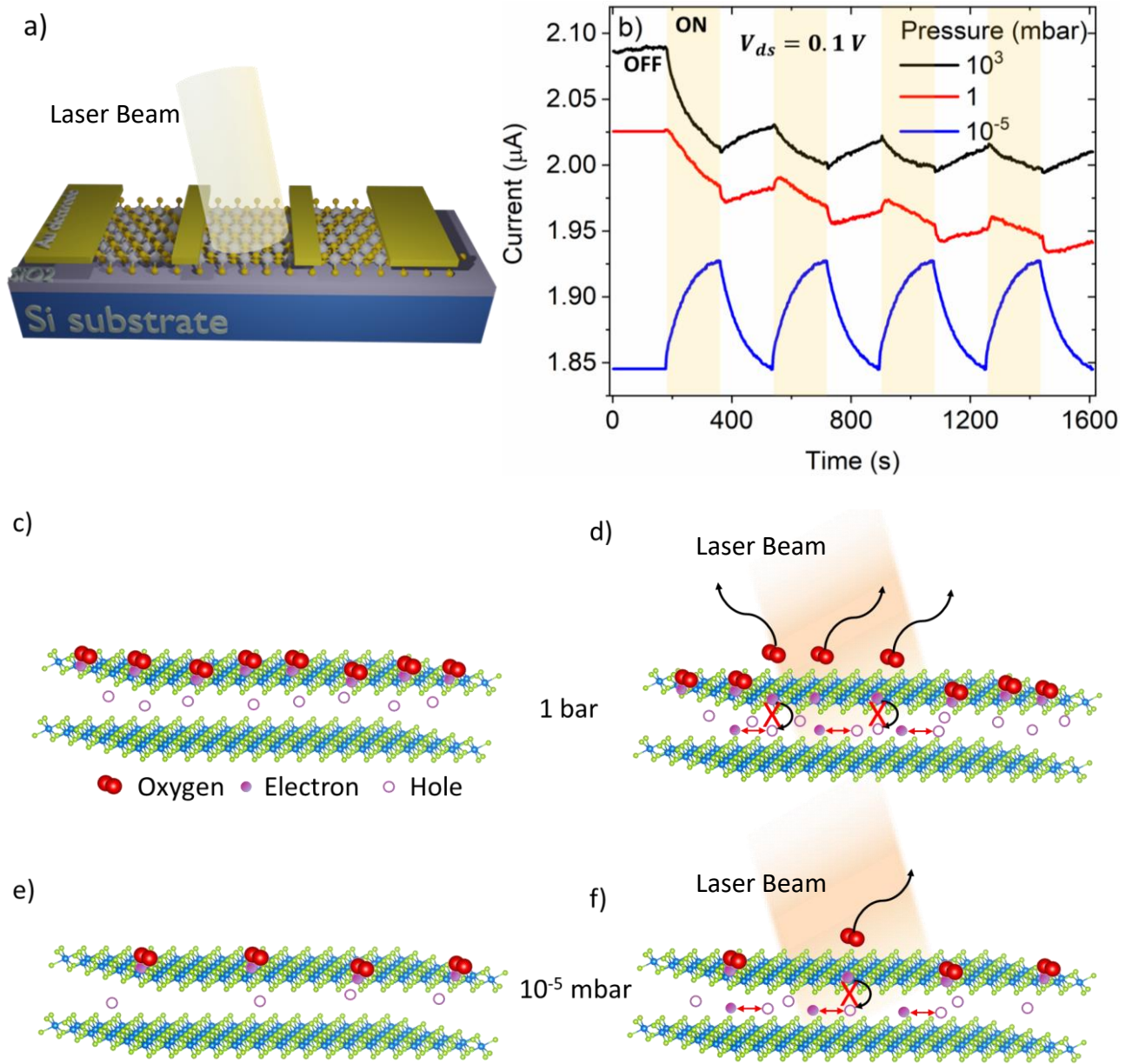

*Figure 3 – Optoelectronic characterization and negative photoconductivity.* *a) Schematic of the device under light irradiation by a super-continuous white laser with wavelength ranging from 450 nm to 2400 nm. The laser spot of about $1\ mm^2$ was located between the two contacts to completely cover the channel. b) Current vs time characteristics at $10^3\ mbar$ (black dots), $1\ mbar$ (red dots), $10^{-5}\ mbar$ (blue dots) under switching light. The laser was switched on (yellow zone) and off (white zone) every 3 minutes. c) At $1\ bar$, in dark, the large amount of oxygen adsorbed on $PtSe_2$ traps electrons at the surface and induces p-type doping in the material. d) When the device is exposed to light, oxygen is removed from the surface. Free*

*electrons are trapped or recombine with holes reducing the density of available carriers and resulting in a negative photoconductivity. e) At $10^{-5}$ mbar the amount of adsorbed oxygen is limited and so is the p-type doping of the material. f) When the device is exposed to light, the photogenerated electron-hole pairs and charge-carrier detrapping increase the conduction in the material, leading to a positive photoconductivity.*

**Figure 3b** shows the device channel current under irradiation, measured at $V_{ds} = 0.1\ V$ and $V_{gs} = 0\ V$. At a pressure of $10^{-5}$ mbar, the dark current is $\sim 1.85\ \mu A$. When the light is turned on, the current increases, reaching a value of $\sim 1.93\ \mu A$. This behaviour is likely dominated by extrinsic photoexcitation, i.e. transitions involving deep trap states, which increase the material conductivity with a characteristic time of $72\ s$ (See Figure S3 in the Supporting Information). First-principles calculations have demonstrated that the formation of deep trap states can be caused by that $Se$ antisites at $Pt$ sites, as well as $Pt$ and $Se$ vacancies.[26] Photoexcitation from deep trap states can result in long response times.[32] Indeed, photocurrent with rising and falling times of dozens seconds have been reported in similar devices with 2D materials such as $MoS_2$[33,34] or $ReS_2$.[32,35]

When the light is switched off, recombination and carrier trapping make the current to return to its initial value. We evaluate the photoresponsivity $R$ of the device through the relation $R = \frac{I_{Light} - I_{Dark}}{PS}$, where $I_{Light}$ and $I_{Dark}$ are the current under illumination and in dark recorded at $V_{ds} = 0.1\ V$ and $V_{gs} = 0\ V$ respectively, $P$ is the incident laser light power and $S$ is the irradiated area. Considering an optical power of $2 \cdot 10^{-8} \frac{W}{\mu m^2}$ and an effective area of $40\ \mu m^2$, we obtained $R \sim 0.2 \frac{A}{W}$. The achieved value is comparable with similar phototransistor realized with both $PtSe_2$ and different 2D materials. Wang et al.[36] reported a low photoresponsivity of $0.19 \frac{mA}{W}$ for $PtSe_2$ phototransistors measured under incident light at $1550\ nm$ and when $V_D =$ $5\ V$, mainly ascribed to insufficient absorption of incident light of the material. Yu et al. showed that such absorption, and thus the photoresponsivity, can be improved by hybridizing $PtSe_2$ with a photonic waveguide.[37] Another route to enhance the photoresponsivity is the use of different substrates. Indeed, Li et al.[38] reported that few-layer $PtSe_2$ on $h-BN$ substrate can reach a photoresponsivity as high as $1.56 \cdot 10^3\ A\ W^{-1}$. Similar studies, conducted on other 2D materials, have demonstrated that the photoconductivity can range from few $\frac{mA}{W}$ from graphene-based devices[39] until to $10^6 \frac{A}{W}$ for black-phosphorus photodetectors.[40] At the pressure of $1\ mbar$, a different behavior is observed. The higher dark current can be ascribed to the interaction between the material surface and the environmental oxygen. When the light is switched on, a sudden increase, appearing as a small step-up, is observed in the current. Such a step is followed by a steady decrease. When the light is switched off, the current suddenly drops with an analogous small step-down and then increases slowly, without attaining its initial value. The successive light pulses result in a further overall decrease of the current. A similar, but more dramatic behaviour is observed at atmospheric pressure. In this case, when the device is illuminated, the current decreases faster and recovers only partially when the light is switched off.

The reduction of current during illumination is known as negative photoconductivity and has been observed in other materials.[41,42] Its origin has not yet been fully understood although most studies attribute it to a photogating effect,[29] an electronic transition to the defect state levels[43] or interaction with adsorbates.[44,45]

The photogating effect is observed when electron-hole photogeneration occurs in $Si$ or $PtSe_2$ followed by charge trapping at the interface with $SiO_2$. Indeed, favoured by the vertical up-bending of the $Si$ bands, photo-generated holes can be trapped at the $Si/SiO_2$ interface acting as a positive gate that lowers the channel conductance.[29] However, this phenomenon should be independent of the external pressure. Since rise and fall time are quite long, the conductance modulation could be ascribed to light induced heating and cooling. Figure 2a shows that the conductivity of the device increases with the temperature independently from the external pressure, hence a thermal effect would not be able to explain the negative photoconductivity at atmospheric pressure. Although the photogating and the thermal effects could play a role, the results of Figure 3b suggest that other mechanisms take place and dominate the optoelectronic behaviour of $PtSe_2$. The dependence on pressure suggests that such mechanisms involve adsorption and desorption of air molecules.[24]

At atmospheric pressure there is a large amount of oxygen molecules adsorbed on the material surface, as shown in **Figure 3c**. Such adsorbates attract free electrons from $PtSe_2$ leaving holes inside the material. These holes are responsible for the p-type conductivity and the higher carrier density measured at atmospheric pressure. When the material is exposed to light, oxygen is easily removed from the surface[46] and the free electrons can recombine with the holes in the material (**Figure 3d**). A similar behaviour has been already reported for $MoS_2$ phototransistors.[47] Indeed, it has been shown that the use of sufficiently short wavelength light source causes desorption of water and oxygen molecules affecting both the responsivity and the response time of the devices. Moreover, although not explicitly shown in Figure 3d, the presence of environmental humidity can also affect the conductivity of the device acting as positive dopants that can be removed through light illumination.[47] The outcome of the process is a decrease in the total concentration of carriers which results in the observed decrease in the conductivity of the material. When the pressure is decreased to $10^{-5}\ mbar$, adsorbed oxygen is partially removed from the surface resulting in the reduced conductivity of the material (**Figure 3e**). In this case, when the channel is exposed to light, the dominant process is the photogeneration of electron-hole pairs and charge detrapping. The excess carriers increase the conductivity of the material, leading to the observed positive photoconductivity (**Figure 3f**). At the intermediate pressure of $1\ mbar$ both processes coexist. Indeed, when the light is first turned on, a step-up in conductivity is observed due to the photogeneration of electron-hole pairs, which is a fast process. Then, the neutralization and desorption of oxygen, which is a slower process, induces a slow decrease of current. When the light is switched off, a step-down in the current is caused by excess electron-hole recombination, followed by a slower current increase due to oxygen re-adsorption on the $PtSe_2$ surface.

## 3. DFT results and discussion

To clarify the impact of physisorbed oxygen molecules on the semiconducting $PtSe_2$ surface, a single oxygen molecule on a bilayer $PtSe_2$ supercell has been considered. Although the experimental film is composed by 6-layers, we performed the DFT calculations on bilayer $PtSe_2$ because of the well-known underestimation of the bandgap in DFT-based calculations.[17] We considered thinner (bilayer) $PtSe_2$ in the DFT calculations to be consistent with the experiments in achieving semiconducting film with comparable bandgap values. As the size of the supercell is larger than the primitive cell, the corresponding first Brillouin zone is smaller. Hence, the $PtSe_2$ slab plus oxygen molecule band structure (supercell structure) is "folded" into the first Brillouin zone. Consequently, to be able to directly compare the folded bands with the reference band structure of the $PtSe_2$ primitive cell, i.e. $PtSe_2$ unit cell without oxygen adsorbate, a procedure known as "unfolding" has been employed to unfold the primitive cell Bloch character hidden in the supercell eigenstates.[48,49] The unfolded band structure of the $PtSe_2$ slab plus oxygen molecule is displayed as a contour plot in **Figure 4a**. The primitive $PtSe_2$ band structure is shown by white dashed lines as reference in this figure and the energies are referenced to the Fermi energy ($E_F$). As can be conspicuously seen by comparing the two band structures, the unfolding method reveals a rather remarkable shift of the $E_F$ towards lower energies, i.e. p-type characteristic, due to the presence of the oxygen adsorbate, while the band structure of the $PtSe_2$ is rather unperturbed.

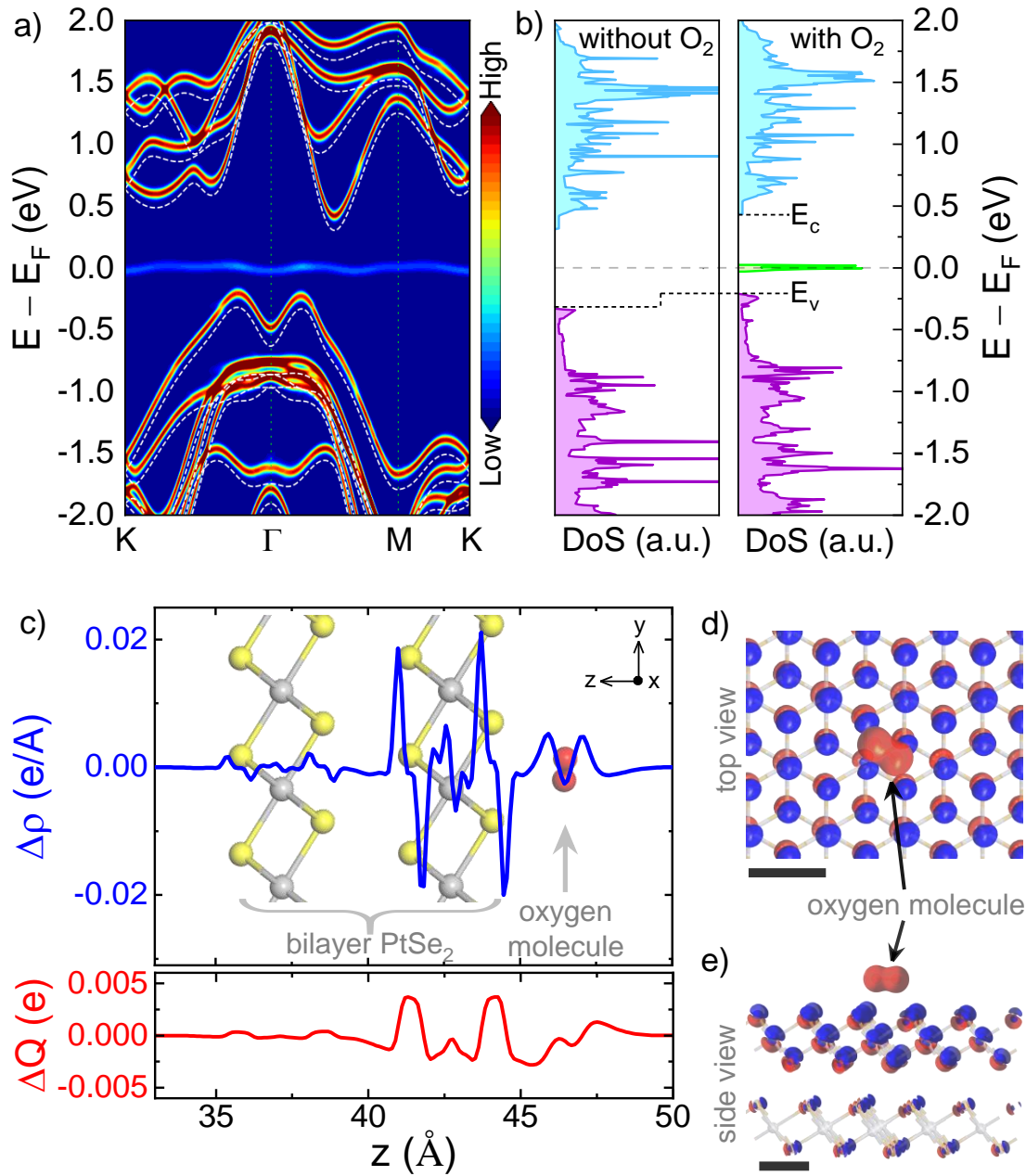

*Figure 4 – **DFT band structure and oxygen effect**. a) Unfolded band structure of the semiconducting PtSe$_2$ with a physisorbed oxygen molecule on the surface using a contour plot of total weight intensity as a function of wave vector in the two-dimensional Brillouin zone. Red (blue) represents high (low) spectral weights. The band structure of the PtSe$_2$ without an oxygen molecule is shown using white dashed lines. Energies are referenced to the Fermi energy. b) DoS of the PtSe$_2$ film with and without oxygen adsorbate on the surface. The Fermi level is at the zero of energy as shown by the grey dashed line. The p-type character of the system with an oxygen molecule is evident. The energy states illustrated in green correspond to*

*the oxygen molecule. c) Plane-averaged charge difference density ($\Delta\rho$) superimposed on the $PtSe_2$ atomic structure with an oxygen molecule, aligned to the z-direction (x-axis) (upper panel in blue) and charge transfer ($\Delta Q$) (lower panel in red) of the structure referenced to the $PtSe_2$ supercell without oxygen molecule. d) and e) Top-view and side-view of the $0.07\ e/Å^3$ CDD isosurface, demonstrating the effects of oxygen adsorbate on the charge distribution in the lateral direction (x-y plane) and along the z-direction, respectively. Red (blue) represents accumulation (depletion) of electrons referenced to the $PtSe_2$ supercell without oxygen molecule. Scale bars in d) and e) are $3.78$ Å.*

Figure 4a also shows the oxygen molecule-induced states are created near the $E_F$ closer to the $PtSe_2$ valence band edge. The nature of these states can be identified through the unfolded band structure, where these states exhibit expected almost dispersionless energy states in the band structure. The density of states (DoS) at the same energy window, with and without the oxygen molecule on the $PtSe_2$ surface, are presented in **Figure 4b**, illustrating the band alignment of the two structures. The occupied double states shown in green are associated with the physisorbed oxygen molecule.

To quantify the charge transfer to/from the molecule, plane-averaged charge difference density (CDD), $\Delta\rho(z)$, along the z-direction is calculated by integrating $\Delta\rho(r)$ within the x-y plane. Accordingly, for a plane located at distance z normal to the $PtSe_2$ plane, the amount of charge transfer is obtained using $\Delta Q(z) = \int_{-\infty}^{z} \Delta\rho(z')\, dz'$. To analyse the charge interaction between the oxygen adsorbate and the semiconducting $PtSe_2$ film, the plane-averaged $\Delta\rho(z)$ referenced to the $PtSe_2$ supercell without oxygen adsorbate, is plotted together with the amount of charge transfer $\Delta Q(z)$ in the upper and lower panel of **Figure 4c**, respectively. As can be seen, the oxygen molecule acts as an electron acceptor by gaining ~0.004 $e$ at the interface from $PtSe_2$ ($e$ is the elementary charge of an electron). The top-view and side-view of the isosurface of the CDD referenced to the $PtSe_2$ supercell without oxygen adsorbate are depicted in **Figure 4d** and **4e**, respectively, where red (blue) represents accumulation (depletion) of electrons. Notably, although the lateral redistribution of charges upon oxygen adsorption is quite negligible (see top-view image in Figure 4d), there is a distinct charge redistribution, particularly between the oxygen molecule and the top layer of the bilayer $PtSe_2$ as can be seen in the side-view (Figure 4e).

## 4. Conclusions

We reported p-type conduction, enhanced by increasing the temperature or pressure, in back-gated field effect transistors with a channel made of ultrathin $PtSe_2$ film. We showed that positive and negative photoconductivity can occur simultaneously and that the dominance of one over the other depends on the environmental pressure. This behaviour was explained by light-induced desorption of oxygen from the surface of the material. Detailed insight into the atomistic effects of a physisorbed oxygen molecule on the band structure of semiconducting $PtSe_2$ has been provided through unfolded band structure calculations. This approach offers a significant advantage of enabling direct comparison of the $PtSe_2$ slab with oxygen adsorbate supercell band structure to the reference $PtSe_2$ primitive cell band structure, revealing oxygen molecule-induced holes in the $PtSe_2$ channel and the shift of the Fermi level towards the $PtSe_2$ valence band edge. The present experimental and theoretical findings can be applied to any other 2D materials with oxygen dependent p-type doping. The understanding of this nontrivial and distinct charge transfer occurring between the physisorbed oxygen molecules and the $PtSe_2$ film provides a potentially effective route for modulating the density of carriers in $PtSe_2$ as ultrathin channel material.

## 5. Experimental Section/Methods

*Fabrication*: For the synthesis of $PtSe_2$, we started from a $Pt$ film (nominal thickness $0.7\ nm$) that was sputtered over an $Si/SiO_2$ substrate ($85\ nm$ thermally grown oxide on p-type silicon, $\rho \sim 0.001 - 0.005 \ \Omega cm$). A two-zone furnace was used for the direct selenization of the $Pt$ film. The upstream zone with $Se$ pellets was heated to 220 °$C$ while the downstream zone housed the $Pt$ film at 400 °$C$ for the selenization process.[21] The obtained $Se$ vapor was carried to the downstream zone through an $Ar:H_2$ (90%:10%) flow at 150 sccm, where it reacted with the $Pt$ films for 2 hours to completely convert it into a $PtSe_2$ ultrathin film. During the process, the $Se$ infiltrated into the $Pt$ film causing the formation of $PtSe_2$ and a film thickness expansion by a factor of $\sim 3.5-4$.[27] Then, the compound was cooled down to room temperature. The final thickness of the $PtSe_2$ film, as revealed by TEM analysis, is about 3 nm corresponding to 6 layers.[26]

The obtained $PtSe_2$ film was transferred to a fresh $Si/SiO_2$ substrate using a polymer-based process and patterned by an $SF_6$-based inductively coupled plasma etching process with photoresist masking. Standard photolithography and lift-off processes were finally applied to define $Ni:Au$ (20 $nm$: 150 $nm$) metal contacts. An optical image of the device is shown in the inset of Figure 1d. AFM image and a schematic of the device are reported in the supporting information.

*Electrical characterizations:* The devices were measured in two- and four-probe configurations in a Janis Probe Station (Janis ST-500 probe station) equipped with four nanoprobes connected to a Keithley 4200 SCS (semiconductor characterization system), working as source-measurement unit with current sensitivity better than 1 pA. The electrical measurements were performed by lowering the pressure from 1 $bar$ (room pressure) to $10^{-5}$ $mbar$ and at different temperatures from 292 K to 392 K. The air in the chamber was evacuated by a rough and a turbo pump. The photoconductivity was tested by irradiating the devices with a super-continuous white laser source (NKT Photonics, Super Compact, wavelength ranging from 450 $nm$ to 2400 $nm$, at 50 $mW/mm^2$).

*First-principles calculations:* First-principles calculations were performed within the framework of density functional theory (DFT) as implemented in QuantumATK.[50] Linear combination of numerical atomic-orbital (LCAO) basis set and generalised gradient approximation (GGA) with norm-conserving pseudopotentials from PseudoDojo[51] were employed in the simulations, where the semi-core 2*s*, 2*p* and 3*d* states for oxygen, 5*s*, 5*p*, 5*d*, 4*f,* 6*s* and 6*p* states for Pt, and 4*s*, 4*p*, 4*f,* and 3*d* states for Se have been retained as valence electrons. Brillouin-zone integrations were performed over a grid of k points generated according to the Monkhorst-Pack scheme[52] with a density of approximately 12 k-points per angstrom. Energy cut-off of 125 Ha has been considered for discretized grid and all structural relaxation was performed with the maximum force of less than $0.005\, eV\, Å^{-1}$.[53] Van der Waals correction to the GGA functional[54] to correct the London dispersion is adopted for analysing the oxygen functionalization of $PtSe_2$. To analyse the effects of the oxygen adsorbate on PtSe$_2$ surface, a single oxygen molecule on a $3 \times 3 \times 1$ bilayer PtSe$_2$ supercell, corresponding to the density of $\sim 9 \times 10^{13}\, cm^{-2}$, has been considered. The slab in the supercell is infinite and periodic in the x- and y-directions (parallel to the slab surface) and is finite along the z-direction (normal to the slab surface). The thickness of the vacuum region along z-direction is larger than 20 Å to avoid any interaction between the periodic images of the neighbouring films. In the slab model calculation with adsorbate on the surface, an artificial macroscopic electrostatic field exists due to the periodic boundary conditions.[55] In order to avoid this artificial field in the $PtSe_2$ slab with oxygen molecule on the surface, we considered mixed Neumann and Dirichlet boundary conditions at the oxygen and $PtSe_2$ sides of the slab, respectively, which provides an alternative approach for the dipole correction in the slab calculations.[50]

**Supporting Information**
Supporting Information is available from the Wiley Online Library or from the author.

**Conflict of interest**
The authors declare no conflict of interest.

**Acknowledgements**

A. Di Bartolomeo acknowledges the funding by the University of Salerno, Salerno, Italy, grants ORSA200207 and ORSA195727 and by the Italian Ministry of University and Research MUR, project RINASCIMENTO ARS01_01088.

Received: ((will be filled in by the editorial staff))
Revised: ((will be filled in by the editorial staff))
Published online: ((will be filled in by the editorial staff))

References

[1] W. Zhang, Y. Zhang, J. Qiu, Z. Zhao, N. Liu, *InfoMat* **2021**, *3*, 133.
[2] S. Manzeli, D. Ovchinnikov, D. Pasquier, O. V. Yazyev, A. Kis, *Nature Reviews Materials* **2017**, *2*, 17033.
[3] B. Radisavljevic, A. Radenovic, J. Brivio, V. Giacometti, A. Kis, *Nature Nanotech* **2011**, *6*, 147.
[4] Y.-Q. Bie, G. Grosso, M. Heuck, M. M. Furchi, Y. Cao, J. Zheng, D. Bunandar, E. Navarro-Moratalla, L. Zhou, D. K. Efetov, T. Taniguchi, K. Watanabe, J. Kong, D. Englund, P. Jarillo-Herrero, *Nature Nanotech* **2017**, *12*, 1124.
[5] S. Wi, H. Kim, M. Chen, H. Nam, L. J. Guo, E. Meyhofer, X. Liang, *ACS Nano* **2014**, *8*, 5270.
[6] A. Pelella, A. Grillo, F. Urban, F. Giubileo, M. Passacantando, E. Pollmann, S. Sleziona, M. Schleberger, A. Di Bartolomeo, *Adv. Electron. Mater.* **2021**, *7*, 2000838.
[7] A. Di Bartolomeo, A. Pelella, F. Urban, A. Grillo, L. Iemmo, M. Passacantando, X. Liu, F. Giubileo, *Adv. Electron. Mater.* **2020**, *6*, 2000094.
[8] F. Liao, Z. Guo, Y. Wang, Y. Xie, S. Zhang, Y. Sheng, H. Tang, Z. Xu, A. Riaud, P. Zhou, J. Wan, M. S. Fuhrer, X. Jiang, D. W. Zhang, Y. Chai, W. Bao, *ACS Appl. Electron. Mater.* **2020**, *2*, 111.
[9] D. J. Late, Y.-K. Huang, B. Liu, J. Acharya, S. N. Shirodkar, J. Luo, A. Yan, D. Charles, U. V. Waghmare, V. P. Dravid, C. N. R. Rao, *ACS Nano* **2013**, *7*, 4879.
[10] F. Urban, F. Giubileo, A. Grillo, L. Iemmo, G. Luongo, M. Passacantando, T. Foller, L. Madauß, E. Pollmann, M. P. Geller, D. Oing, M. Schleberger, A. Di Bartolomeo, *2D Mater.* **2019**, *6*, 045049.
[11] E. Chen, W. Xu, J. Chen, J. H. Warner, *Materials Today Advances* **2020**, *7*, 100076.
[12] A. Avsar, A. Ciarrocchi, M. Pizzochero, D. Unuchek, O. V. Yazyev, A. Kis, *Nat. Nanotechnol.* **2019**, *14*, 674.
[13] Y. Zhao, J. Qiao, Z. Yu, P. Yu, K. Xu, S. P. Lau, W. Zhou, Z. Liu, X. Wang, W. Ji, Y. Chai, *Adv. Mater.* **2017**, *29*, 1604230.
[14] G. Y. Guo, W. Y. Liang, *J. Phys. C: Solid State Phys.* **1986**, *19*, 995.
[15] H. L. Zhuang, R. G. Hennig, *J. Phys. Chem. C* **2013**, *117*, 20440.
[16] Y. Wang, L. Li, W. Yao, S. Song, J. T. Sun, J. Pan, X. Ren, C. Li, E. Okunishi, Y.-Q. Wang, E. Wang, Y. Shao, Y. Y. Zhang, H. Yang, E. F. Schwier, H. Iwasawa, K. Shimada, M. Taniguchi, Z. Cheng, S. Zhou, S. Du, S. J. Pennycook, S. T. Pantelides, H.-J. Gao, *Nano Lett.* **2015**, *15*, 4013.
[17] L. Ansari, S. Monaghan, N. McEvoy, C. Ó. Coileáin, C. P. Cullen, J. Lin, R. Siris, T. Stimpel-Lindner, K. F. Burke, G. Mirabelli, R. Duffy, E. Caruso, R. E. Nagle, G. S. Duesberg, P. K. Hurley, F. Gity, *npj 2D Mater Appl* **2019**, *3*, 33.
[18] W. Zhang, J. Qin, Z. Huang, W. Zhang, *Journal of Applied Physics* **2017**, *122*, 205701.

[19] X. Chia, A. Adriano, P. Lazar, Z. Sofer, J. Luxa, M. Pumera, *Adv. Funct. Mater.* **2016**, *26*, 4306.
[20] T.-Y. Su, H. Medina, Y.-Z. Chen, S.-W. Wang, S.-S. Lee, Y.-C. Shih, C.-W. Chen, H.-C. Kuo, F.-C. Chuang, Y.-L. Chueh, *Small* **2018**, *14*, 1800032.
[21] C. Yim, K. Lee, N. McEvoy, M. O'Brien, S. Riazimehr, N. C. Berner, C. P. Cullen, J. Kotakoski, J. C. Meyer, M. C. Lemme, G. S. Duesberg, *ACS Nano* **2016**, *10*, 9550.
[22] M. Sajjad, E. Montes, N. Singh, U. Schwingenschlögl, *Adv. Mater. Interfaces* **2017**, *4*, 1600911.
[23] T.-Y. Su, Y.-Z. Chen, Y.-C. Wang, S.-Y. Tang, Y.-C. Shih, F. Cheng, Z. M. Wang, H.-N. Lin, Y.-L. Chueh, *J. Mater. Chem. C* **2020**, *8*, 4851.
[24] Q. Liang, J. Gou, Arramel, Q. Zhang, W. Zhang, A. T. S. Wee, *Nano Res.* **2020**, *13*, 3439.
[25] Y. Gong, Z. Lin, Y.-X. Chen, Q. Khan, C. Wang, B. Zhang, G. Nie, N. Xie, D. Li, *Nano-Micro Lett.* **2020**, *12*, 174.
[26] H. Zheng, Y. Choi, F. Baniasadi, D. Hu, L. Jiao, K. Park, C. Tao, *2D Mater.* **2019**, *6*, 041005.
[27] C. Yim, V. Passi, M. C. Lemme, G. S. Duesberg, C. Ó Coileáin, E. Pallecchi, D. Fadil, N. McEvoy, *npj 2D Mater Appl* **2018**, *2*, 5.
[28] Z.-X. Zhang, Long-Hui Zeng, X.-W. Tong, Y. Gao, C. Xie, Y. H. Tsang, L.-B. Luo, Y.-C. Wu, *J. Phys. Chem. Lett.* **2018**, *9*, 1185.
[29] F. Urban, F. Gity, P. K. Hurley, N. McEvoy, A. Di Bartolomeo, *Appl. Phys. Lett.* **2020**, *117*, 193102.
[30] F. Giannazzo, G. Fisichella, A. Piazza, S. Di Franco, G. Greco, S. Agnello, F. Roccaforte, *Beilstein J. Nanotechnol.* **2017**, *8*, 254.
[31] S. M. Sze, K. K. Ng, *Physics of semiconductor devices*, 3rd ed., Wiley-Interscience, Hoboken, N.J, **2007**.
[32] J. Jiang, C. Ling, T. Xu, W. Wang, X. Niu, A. Zafar, Z. Yan, X. Wang, Y. You, L. Sun, J. Lu, J. Wang, Z. Ni, *Adv. Mater.* **2018**, *30*, 1804332.
[33] W.-C. Shen, R.-S. Chen, Y.-S. Huang, *Nanoscale Res Lett* **2016**, *11*, 124.
[34] M.-A. Kang, S. Kim, I.-S. Jeon, Y. R. Lim, C.-Y. Park, W. Song, S. S. Lee, J. Lim, K.-S. An, S. Myung, *RSC Adv.* **2019**, *9*, 19707.
[35] E. Liu, M. Long, J. Zeng, W. Luo, Y. Wang, Y. Pan, W. Zhou, B. Wang, W. Hu, Z. Ni, Y. You, X. Zhang, S. Qin, Y. Shi, K. Watanabe, T. Taniguchi, H. Yuan, H. Y. Hwang, Y. Cui, F. Miao, D. Xing, *Adv. Funct. Mater.* **2016**, *26*, 1938.
[36] Y. Wang, Z. Yu, Y. Tong, B. Sun, Z. Zhang, J.-B. Xu, X. Sun, H. K. Tsang, *Appl. Phys. Lett.* **2020**, *116*, 211101.
[37] Z. Yu, Y. Wang, B. Sun, Y. Tong, J. Xu, H. K. Tsang, X. Sun, *Adv. Optical Mater.* **2019**, *7*, 1901306.
[38] L. Li, W. Wang, Y. Chai, H. Li, M. Tian, T. Zhai, *Adv. Funct. Mater.* **2017**, *27*, 1701011.
[39] M. Furchi, A. Urich, A. Pospischil, G. Lilley, K. Unterrainer, H. Detz, P. Klang, A. M. Andrews, W. Schrenk, G. Strasser, T. Mueller, *Nano Lett.* **2012**, *12*, 2773.
[40] M. Huang, M. Wang, C. Chen, Z. Ma, X. Li, J. Han, Y. Wu, *Adv. Mater.* **2016**, *28*, 3481.
[41] G. Z. Liu, R. Zhao, J. Qiu, Y. C. Jiang, J. Gao, *J. Phys. D: Appl. Phys.* **2019**, *52*, 095302.
[42] Y. Zhang, X. Li, X. Lin, G. Li, Y. Cai, C. Wen, K. Wang, D. Liu, S. Hu, Y. Hu, *Materials Science in Semiconductor Processing* **2019**, *98*, 106.
[43] X. Xiao, J. Li, J. Wu, D. Lu, C. Tang, *Appl. Phys. A* **2019**, *125*, 765.
[44] Y. Han, X. Zheng, M. Fu, D. Pan, X. Li, Y. Guo, J. Zhao, Q. Chen, *Phys. Chem. Chem. Phys.* **2016**, *18*, 818.
[45] Y. Liu, P. Fu, Y. Yin, Y. Peng, W. Yang, G. Zhao, W. Wang, W. Zhou, D. Tang, *Nanoscale Res Lett* **2019**, *14*, 144.
[46] Y. Wang, Z. He, J. Zhang, H. Liu, X. Lai, B. Liu, Y. Chen, F. Wang, L. Zhang, *Nano Res.* **2020**, *13*, 358.

[47] P. Han, E. R. Adler, Y. Liu, L. St Marie, A. El Fatimy, S. Melis, E. Van Keuren, P. Barbara, *Nanotechnology* **2019**, *30*, 284004.
[48] C.-C. Lee, Y. Yamada-Takamura, T. Ozaki, *J. Phys.: Condens. Matter* **2013**, *25*, 345501.
[49] F. Gity, F. Meaney, A. Curran, P. K. Hurley, S. Fahy, R. Duffy, L. Ansari, *Journal of Applied Physics* **2021**, *129*, 015701.
[50] S. Smidstrup, T. Markussen, P. Vancraeyveld, J. Wellendorff, J. Schneider, T. Gunst, B. Verstichel, D. Stradi, P. A. Khomyakov, U. G. Vej-Hansen, M.-E. Lee, S. T. Chill, F. Rasmussen, G. Penazzi, F. Corsetti, A. Ojanperä, K. Jensen, M. L. N. Palsgaard, U. Martinez, A. Blom, M. Brandbyge, K. Stokbro, *J. Phys.: Condens. Matter* **2020**, *32*, 015901.
[51] M. J. van Setten, M. Giantomassi, E. Bousquet, M. J. Verstraete, D. R. Hamann, X. Gonze, G.-M. Rignanese, *Computer Physics Communications* **2018**, *226*, 39.
[52] H. J. Monkhorst, J. D. Pack, *Phys. Rev. B* **1976**, *13*, 5188.
[53] L. Ansari, G. Fagas, F. Gity, J. C. Greer, *Appl. Phys. Lett.* **2016**, *109*, 063108.
[54] M. Dion, H. Rydberg, E. Schröder, D. C. Langreth, B. I. Lundqvist, *Phys. Rev. Lett.* **2004**, *92*, 246401.
[55] J. Neugebauer, M. Scheffler, *Phys. Rev. B* **1992**, *46*, 16067.